\documentclass{Interspeech2024}

\interspeechcameraready

\title{Pretraining End-to-End Keyword Search\\ with Automatically Discovered Acoustic Units}

\name[affiliation={1,2}]{Bolaji}{Yusuf}
\name[affiliation={2}]{Jan}{``Honza" Cernocky}
\name[affiliation={1}]{Murat}{Saraclar}

\address{
  $^1$Bogazici University, Turkey \\
  $^2$Brno University of Technology, Speech@FIT, Czechia
  }
\email{\{iyusuf,cernocky\}@fit.vut.cz, murat.saraclar@bogazici.edu.tr}

\usepackage{algorithm2e}
\usepackage{amsmath,graphicx}
\usepackage{amssymb}
\usepackage{array}
\usepackage{bbm}
\usepackage{booktabs}
\usepackage{cite}
\usepackage[utf8]{inputenc}
\usepackage{mathtools}
\usepackage{mystyle}
\usepackage{multirow}
\usepackage{tikz}
\usepackage{hyperref}
\usepackage{pgfplots}
\usepackage{stfloats}
\usepackage{pifont}

\usepackage{subcaption,caption}
\usepackage{textcomp}
\usepackage{tikz-3dplot}
\usepackage{listings,xcolor}
\usepackage{url}
\usepackage{verbatim}
\usetikzlibrary{arrows.meta}
\usetikzlibrary{positioning,through, backgrounds,decorations.pathreplacing,shapes.misc}
\pgfplotsset{compat=1.17}
\pgfplotsset{every axis/.append style={
                    label style={font=\Large},
                    tick label style={font=\Large}  
                    }}

\tdplotsetmaincoords{70}{30}

\definecolor{codegreen}{rgb}{0,0.6,0}
\definecolor{codegray}{rgb}{0.5,0.5,0.5}
\definecolor{codepurple}{rgb}{0.58,0,0.82}
\definecolor{backcolour}{rgb}{0.95,0.95,0.92}
\lstdefinestyle{lsty}{
    backgroundcolor=\color{backcolour},   
    commentstyle=\color{codegreen},
    keywordstyle=\color{magenta},
    numberstyle=\tiny\color{codegray},
    stringstyle=\color{codepurple},
    basicstyle=\ttfamily\footnotesize,
    breakatwhitespace=false,         
    breaklines=true,                 
    captionpos=b,                    
    keepspaces=true,                 
    numbers=left,                    
    numbersep=5pt,                  
    showspaces=false,                
    showstringspaces=false,
    showtabs=false,                  
    tabsize=1
}
\begin{document}

\maketitle

\begin{abstract}
   End-to-end (E2E) keyword search (KWS) has emerged as an alternative and complimentary approach to conventional keyword search which depends on the output of automatic speech recognition (ASR) systems.
   While E2E methods greatly simplify the KWS pipeline, they generally have worse performance than their ASR-based counterparts, which can benefit from pretraining with untranscribed data.
   In this work, we propose a method for pretraining E2E KWS systems with untranscribed data, which involves using acoustic unit discovery (AUD) to obtain discrete units for untranscribed data and then learning to locate sequences of such units in the speech.
   We conduct experiments across languages and AUD systems: we show that finetuning such a model significantly outperforms a model trained from scratch, and the performance improvements are generally correlated with the quality of the AUD system used for pretraining.
\end{abstract}

\section{Introduction}
Productive use of the internet relies heavily on the presence and capacity of search engines to efficiently index and search through large quantities of data.
Since a significant proportion of that data is in multimedia form, it is natural to develop technologies to allow efficient search through non-textual documents.
Keyword search (KWS), also known as spoken term detection (STD), is one such technology: it aims to locate where in an archive of spoken documents a user-specified query has been uttered.
A KWS system takes a written query and returns a list of documents purported to contain the query, timestamps in those documents where the query is located, and scores representing the system's confidence in its hypotheses.

KWS is traditionally done by conducting text-based retrieval on the output of an automatic speech recognition (ASR) system.
Outside settings with very-low recognition error rates where one-best ASR outputs may be sufficient, it is more common to index richer structures like lattices or confusion networks~\cite{Ng2000,szoke2008hybrid,can2011lattice,chelba2008retrieval,mangu2014efficient}, which improve recall by accounting for uncertainty in ASR output.

More recently, ASR-free KWS methods have sought to eschew the ASR and its concomitant complexities~\cite{audhkhasi2017end,gundougdu2017joint,yusuf2019low,yusuf21_interspeech,svec21_interspeech,fuchs2021cnn,yusuf2023end}.
Instead of relying on the output of an ASR system\footnote{This refers to indexing and search. Even ASR-free KWS systems generally rely on simple ASR systems to get timing information required for training.}, a neural network is trained in an end-to-end (E2E) fashion to locate written queries in large spoken archives.
We take~\cite{yusuf2023end} as a representative of this approach, and use it as our baseline in this work.
The KWS model comprises a pair of encoders: a query encoder that takes a query in the form of a sequence of letters and computes a vector representation thereof, and a document encoder for computing a compatible representation of the spoken document.
The two are combined via frame-wise inner-products and locations in the document which have high inner-products with the query embedding are returned as hits.

Although E2E KWS methods are able to streamline the indexing and search, they generally trail ASR-based methods in terms of search accuracy.
Furthermore, ASR-based systems can benefit from the rise in semi-supervised learning that improve the underlying ASR model by using large amounts of untranscribed speech for pretraining.
Our objective in this paper is, therefore, to design a pretraining scheme for E2E KWS to be able to leverage untranscribed speech.
We note here that~\cite{yusuf2023end} already explored pretraining for E2E-KWS, but it only considered pretraining with transcribed multilingual data, while we explore pretraining with untranscribed data in the target language---with the potential to expand to multilingual untranscribed data.

The input data for training E2E KWS comprises sets of speech documents and the words (queries) they contain.
Hence, to pretrain E2E KWS on an untranscribed speech corpus with the same training objective, the challenge is to get the list of queries which we can pretrain the KWS model to locate.
In other words, we need sequences of discrete units corresponding] to the spoken content.

Acoustic unit discovery (AUD) aims to solve this exact problem---automatically discovering an inventory of phone-like units for a language from completely unlabeled data.
Several works have tackled AUD, including Bayesian methods~\cite{lee2012nonparametric,chen2015parallel,heck2016iterative,ondel2016variational}, neural-network-based methods~\cite{chorowski2019unsupervised,Baevski2020vq-wav2vec,Harwath2020Learning} or hybrids thereof~\cite{ebbers2017hidden,glarner2018full}.

We employ the Hierarchical Subspace Hidden Markov Model (H-SHMM)~\cite{yusuf2021hierarchical,ondel2022nonparametric}, a non-parametric Bayesian model for AUD.
H-SHMM models follows the phone-loop AUD paradigm~\cite{lee2012nonparametric,ondel2016variational,Ondel2019shmm}, where each acoustic unit is modeled as an HMM.
In H-SHMM, the HMM parameters are constrained to a phonetic subspace of the parameter space and the parameters that define the subspace itself are allowed to vary per language within a constrained ``hyper"-subspace.
We choose H-SHMM as it has good performance not just on intrinsic AUD but also unsupervised word segmentation~\cite{boito2022unsupervised}, which makes it suitable for our task of word localization.
After training the H-SHMM, we use it to label the untranscribed speech.
Then we use sequences of acoustic unit labels as pseudo-queries for pretraining the model.
Finally, we finetune the model on a small amount of transcribed data.

We conduct experiments on the English Libri-light~\cite{librilight} and Turkish Broadcast News~\cite{arisoy2009turkish} corpora.
Our experiments show that AUD-based pretraining significantly improves KWS performance for AUD that uses MFCC features, with improvements that are correlated with the AUD system's phonetic correspondence.
Furthermore, when the AUD uses pretrained transformer features as input, we still get significant improvements on Libri-light.

\section{Methods}
\begin{figure}[t]
    \centering
    \begin{subfigure}[b]{0.48\linewidth}
        \centering
        \includegraphics[width=\linewidth]{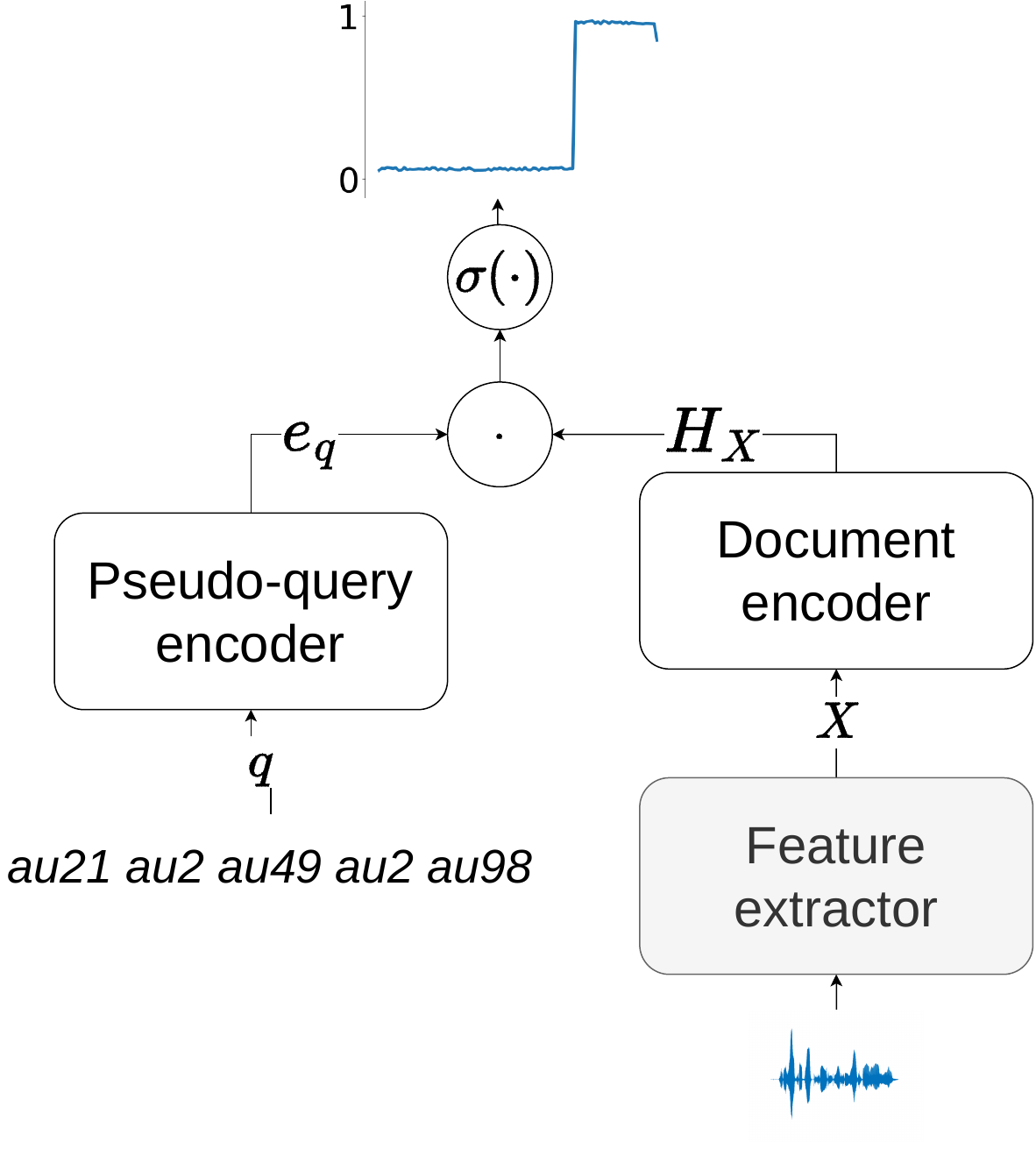}
        \caption{Pretraining.}
        \label{fig:baseline}
    \end{subfigure}
    ~
    \begin{subfigure}[b]{0.48\linewidth}
        \centering
        \includegraphics[width=\linewidth]{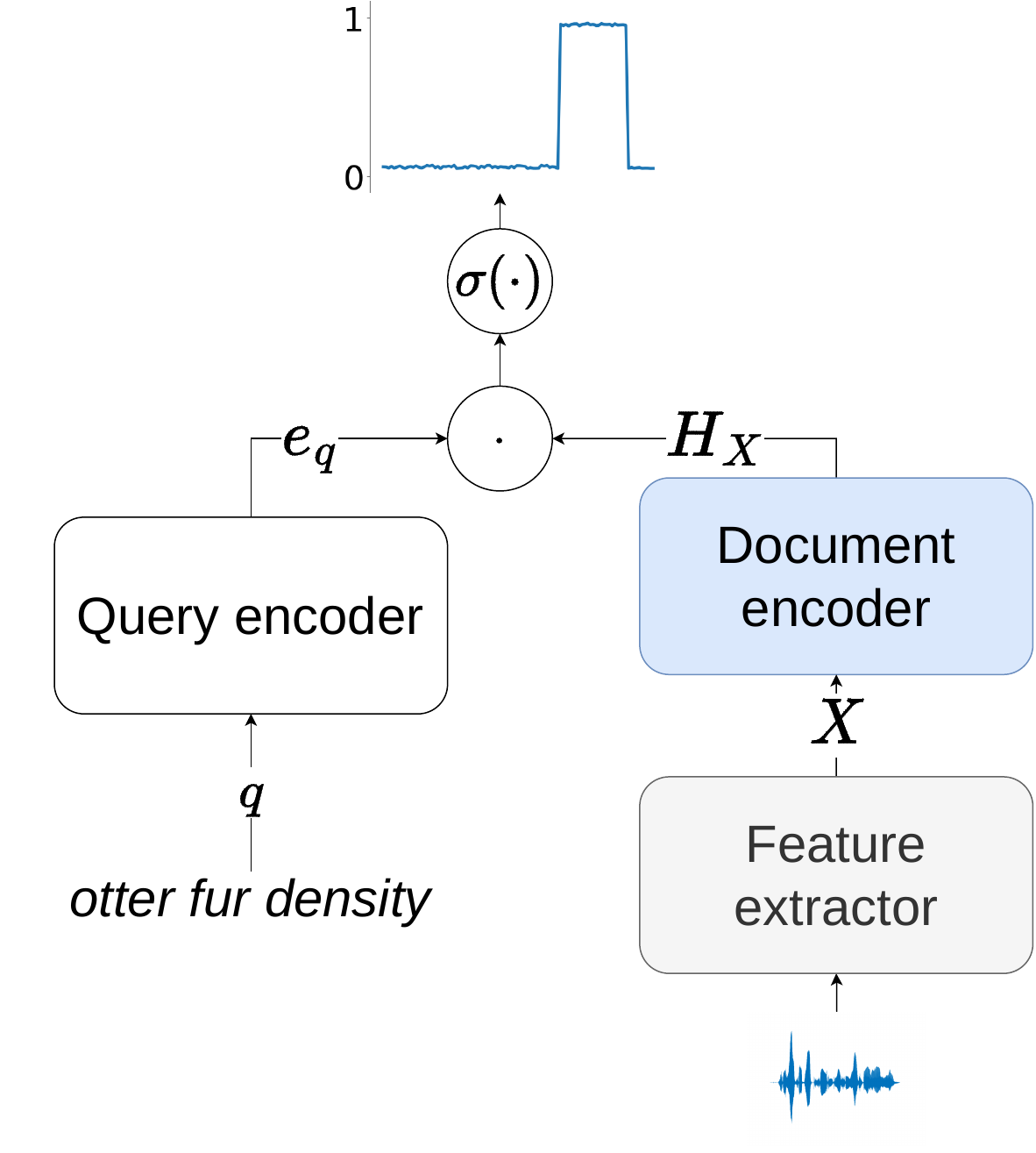}
        \caption{Finetuning.}
        \label{fig:proposed}
    \end{subfigure}
    \caption{E2E KWS system during pretraining and finetuning.
    Grey boxes: non-trainable components, white boxes: components which are trained from scratch, blue boxes: trainable components whose parameters are transferred from a pretrained model.
    }
    \label{fig:model}
\end{figure}

\subsection{Background}
\subsubsection{End-to-end Keyword Search}
\label{sec:methods:e2ekws}
Our work is based on the E2E KWS model of~\cite{yusuf2023end}.
This model ingests a textual query in the form of a sequence of $L$ letters, $\Matrix{q} = (q_1, \dots, q_L)$, and a spoken utterance of $N$ frames, $\Matrix{X} = (\Matrix{x}_1, \dots, \Matrix{x}_N)$ and predicts the sequence $\Matrix{y}(\Matrix{q}, \Matrix{X}) = (y_1, \dots, y_{N})$, where each $y_n \in \{0, 1\}$ is a binary random variable indicating the existence of the query in the $n$th frame of the document, i.e:
\begin{align}
    y_n=
    \begin{cases}
      1, & \text{if}\ \Matrix{q} \text{ is spoken in } \Matrix{X} \text{ in a time span including } n \\
      0, & \text{otherwise}.
    \end{cases}
    \label{eqn:y_speech}
\end{align}
The model comprises a pair of encoders:
\begin{itemize}
    \item The query encoder computes a fixed-length representation, $\Matrix{e}_{\Matrix{q}}$, of the query by passing it through a GRU, summing the activations from the last layer of the GRU across the sequence dimension, followed by an affine projection.
    \item The document encoder computes a down-sampled representation, of the document $\Matrix{H}_{\Matrix{X}} = (\Matrix{h}_1, \dots, \Matrix{h}_N)$ by passing it through a BLSTM followed by an affine transform.
\end{itemize}
The sequence $\Matrix{z}(\Matrix{q}, \Matrix{X}) = (z_1, \dots, z_N)$ of occurrence probabilities, $z_n \coloneqq P(y_n = 1 | \Matrix{q}, \Matrix{X})$, is then obtained via a matrix-vector product of $\Matrix{H}_{\Matrix{X}}$ and $\Matrix{e}_{\Matrix{q}}$ as:
\begin{align}
    z_n = \sigma(\Matrix{h}_n ^\top \Matrix{e}_{\Matrix{q}}),
    \label{eq:output}
\end{align}
where $\sigma(\cdot)$ is the logistic sigmoid function.

The vector of probabilities from~\eqref{eq:output} is then post-processed to obtain the timestamps in the document hypothesized to contain the query, and the corresponding confidence scores by detecting ``islands" of probabilities above $0.5$.
The procedure is as follows:
\begin{enumerate}
    \item Probabilities $z_n < 0.5$ are zeroed-out.
    \item The resulting ``islands" of non-zero elements are returned as system hypotheses, and each hypothesis' confidence score is computed as the median probability in its time-span.
\end{enumerate}

The model is trained with mini-batch gradient descent on a transcribed speech dataset.
At each training step, $t$, the following objective is minimized:
\begin{align}
    J_t = \sum_{l=1}^{L} \sum_{m=1}^M f \Bigl(
        & \Matrix{z} \bigl(\Matrix{q}_{t, l}, \Matrix{X}_{t,l,m}\bigr), \Matrix{y} \bigl(\Matrix{q}_{t, l}, \Matrix{X}_{t,l,m}\bigr)
    \Bigr),
    \label{eqn:single_step}
\end{align}
where $\{\Matrix{q}_{t, 1} \dots \Matrix{q}_{t, L} \}$ is a mini-batch of $L$ queries sampled randomly from the set of unigrams, bigrams and trigrams of the dataset;
$\{\Matrix{X}_{t,l,1}, \dots, \Matrix{X}_{t,l,M}\}$ is a set of documents sampled from the dataset such that $\{\Matrix{X}_{t,l,1}\}$ contains $\Matrix{q}_{t, l}$ while the other $M-1$ documents are sampled randomly;
and $f(\cdot)$ is a modified binary cross-entropy function defined as:
\begin{align}
    f(z, y) = -\sum_{n=1}^{\hat{N}}& \Bigl (\mathbbm{1}_{z_n > 1 - \phi} \cdot (1-y_{n}) \log (1-z_n)
     \nonumber \\
    & + \mathbbm{1}_{z_n < \phi} \cdot \lambda \cdot y_{n} \log z_n \Bigr),
\end{align}
with $\phi$ controlling the tolerance of the objective to easily-classified frames and $\lambda$ controlling the relative weighting of positive to negative frames.
Following~\cite{yusuf2023end}, we set $\lambda=5$, $\phi=0.7$ and $M=4$ in all our experiments.

The word-level alignments required for training are obtained by training an HMM-GMM ASR system on the training data and using it for forced alignment.

\subsubsection{Hierarchical Subspace Hidden Markov Model}
\label{sec:methods:hshmm}
Acoustic unit discovery entails learning a set of units from untranscribed data.
For a language, $l$, this typically involves learning a set of parameters, $\Theta^l = \{ \Matrix{\theta}^{l, u} \}_{u=1}^{U_l}$ for each unit $u$, which then allows frames of an utterance $\Matrix {X} = (\Matrix{x}_1, \dots, \Matrix{x}_n)$ in that language to be labeled into discrete units $v_1, \dots, v_N$ where each $v_n \in \{1, 2, \dots, U_l\}$.

We employ H-SHMM~\cite{yusuf2021hierarchical}, a Dirichlet-process-based Bayesian nonparametric model for AUD.
In H-SHMM, each acoustic unit is a 3-state, left-to-right HMM-GMM with 4 Gaussians per state, and each $\Matrix{\theta}^{l,u}$ is a super-vector formed by concatenating all the mean vectors, covariance matrix elements and mixture weights of the HMM-GMM.
The parameters of the HMM-GMMs are constrained to dwell in a low-dimensional subspace of the full parameter space:
\begin{align}
    \Matrix{\theta}^{l,u} = g \bigl( \Matrix{W}^l \Matrix{\eta}^{l,u} + \Matrix{b}^l \bigr),
\end{align}
where $\Matrix{\eta}^{l,u}$ is a low-dimensional\footnote{We set the dimensionality of $\Matrix{\eta}^{l,u}$ to 100 in our experiments. Contrast this to the dimensionality of $\Matrix{\theta}^{l,u}$ which is around 1000 for HMMs with MFCC features and around 25000 for HMMs with XLS-R features.} embedding of the parameter, $\Matrix{W}^l$ is a language-specific low-rank matrix which, along with the bias vector $\Matrix{b}^l$, defines the subspace to which the parameters are constrained, and $g(\cdot)$ is a non-linear function mapping vectors from the column space of $\Matrix{W}^l$ to HMM parameters, ensuring e.g. that the dimensions corresponding to each covariance matrix constitute a positive-definite matrix and that the mixture weights are non-negative and sum up to one.
The subspace parameters are themselves further constrained to a $K$-dimensional ``hyper"-subspace:
\begin{align}
    \Matrix{W}^l = \Matrix{M}_0 + \sum_{k=1} ^K \alpha^{l}_k + \Matrix{M}_k \nonumber \\
    \Matrix{b}^l = \Matrix{m}_0 + \sum_{k=1} ^K \alpha^{l}_k + \Matrix{m}_k,
\end{align}
where $\Matrix{\alpha}^l$ is a language-specific low-dimensional embedding, and $\{\Matrix{M}_k \}$ and $\{\Matrix{m}_k \}$ define language-agnostic template matrices and vectors whose linear combinations define the part of the space $\Matrix{W}^l$ and $\Matrix{b}^l$ are allowed to occupy.
Thus, the subspace parameters are allowed to vary in a controlled manner from one language to another.
The H-SHMM defines a distribution with trainable parameters---$\{\Matrix{M}_k\}, \{\Matrix{m}_k\}, \{\Matrix{\alpha}^l\}, \{\Matrix{\eta}^{l, u}\}$---which are learned in two phases:
\begin{enumerate}
    \item Supervised pretraining: The model is first trained on phonetically-transcribed speech from multiple languages not including the target language.
    This imbues the templates $\{\Matrix{M}_k \}$ and $\{\Matrix{m}_k \}$ with phone-like characteristics.
    \item Acoustic unit discovery: The distributions $\{\Matrix{M}_k \}$ and $\{\Matrix{m}_k \}$ are kept fixed, and transferred to a target language, $l^*$, for which $\Matrix{\alpha}^{l^*}$ and $\{\Matrix{\eta}^{l^*, u}\}$ are learned.
\end{enumerate}

Both phases involve optimizing a variational lower bound on the log-likelihood of the data, which yields a Baum-Welch-like training procedure.
Having obtained the distributions of the HMM parameters, the untranscribed speech can them be labelled with a variational analog of Viterbi decoding.
Interested readers are referred to~\cite{ondel2022nonparametric} for a thorough coverage of H-SMMM and its inference.

\subsection{Pretraining KWS with AUD}
\label{sec:methods:pretraining}
In this paper, we propose using AUD to label an otherwise untranscribed speech corpus into acoustic units, and using these pseudo-labels to pretrain the E2E KWS model in a setting where we have a small transcribed corpus and a larger untranscribed speech corpus in the same language.

We train an H-SHMM on a small subset of the unlabeled speech and use it to transcribe the full corpus into sequences of acoustic units.
Since the KWS model expects word sequences as input, and the AUD only returns phone-like units, we form pseudo-words from acoustic unit n-grams.
Specifically, we take all sequences of 5 to 15 consecutive acoustic units as pseudo-words, and use these pseudo-words as queries to the KWS model, which we pretrain using~\eqref{eqn:single_step} to locate them in the corpus.
Note that we have the pseudo-word time boundaries for training since decoding with H-SHMM, as with any other HMM, naturally yields frame-level decisions for acoustic units.

After pretraining is complete, we transfer the document encoder and discard the acoustic unit query encoder.
We then initialize a new query encoder for actual graphemes and train it along with the transferred document encoder to locate real queries as described in Section~\ref{sec:methods:e2ekws}.

\section{Experiments}
\subsection{Setup}
\subsubsection{Datasets}
\label{sec:experiments:datasets}
\noindent \textbf{Keyword search}:
We test the performance on the Libri-light~\cite{librilight} and Turkish Broadcast News (BNTR)~\cite{arisoy2009turkish}{} corpora.
For Libri-light, we use the 10 hour Libri-light training set as the transcribed data and test on the standard Librispeech~\cite{panayotov2015librispeech} sets (dev-clean, test-clean, dev-other and dev-other).
To match the Libri-light training data size, we use a 10-hour subset of BNTR from the VOA programs\footnote{https://catalog.ldc.upenn.edu/LDC2012S06} for training and select two 10-hour subsets from the remaining BNTR data as dev and test sets.
Since neither dataset has official query lists, we randomly select 1500 queries composed of equal proportions of unigrams, bigrams and trigrams for each of the dev and eval sets.
Table~\ref{tab:text_stats} shows the proportion of out-of-vocabulary (OOV) queries.

For Libri-light pretraining, we use the 360-hour set of Librispeech as untranscribed data.
In the case of BNTR, we use the full 180-hour BNTR training set for pretraining.
\begin{table}[t]
    \centering
    \caption{OOV rates for the query lists used for each dev/test set.}
    \label{tab:text_stats}
    \centering
    \begin{tabular}{l cc ccc ccc c|cc ccc ccc ccc}
    \toprule
          Dataset & \multicolumn{2}{c}{LS-clean} & \multicolumn{2}{c}{LS-other} & \multicolumn{2}{c}{BNTR} \\
         & Dev & Test & Dev & Test & Dev & Test \\
         \midrule
         OOV-Rate (\%)& 1.1 & 2.6 & 2.5 & 3.6 & 11.9 & 6.3 \\
         \bottomrule
    \end{tabular}
\end{table}

\noindent \textbf{Acoustic unit discovery}: For the supervised phase of H-SHMM training where we estimate the hyper-subspaces ($\{\Matrix{M}_k\}$, and $\{\Matrix{m}_k\}$ in Section~\ref{sec:methods:hshmm}), we use the models from~\cite{ondel2022nonparametric}, trained on data from seven languages\footnote{https://github.com/beer-asr/beer/tree/master/recipes/hshmm}: French, German, Polish and Spanish from the Globalphone corpus~\cite{schultz2013globalphone}, and Amharic, Swahili and Wolof from the Alffa project~\cite{besacier2015speech}.
A 1500-utterance subset of each language (totalling around 19 hours of speech) was used.

For actual acoustic unit discovery, where we learn the acoustic units ($\{\Matrix{\alpha}^l\}, \{\Matrix{\eta}^{l, u}\}$) for each target language, we use random 3000-utterance subsets from each respective language's untranscribed data, and use the learned HMMs to transcribe the full corpus.

\subsubsection{Acoustic features}
The default acoustic inputs to our models (both AUD and KWS) are 13-dimensional MFCC features.
In addition, we consider features from a pretrained 300 million parameter XLS-R model~\cite{babu22_interspeech}\footnote{https://dl.fbaipublicfiles.com/fairseq/wav2vec/xlsr2\_300m.pt}, from which we use the output of the 15th layer, shown in~\cite{ondel2022nonparametric} to yield considerably better AUD performance.
Note that, due to computational constraints, we only use the XLS-R model as a feature extractor and we do not finetune it.

\begin{table*}[t]
    \caption{Term weighted value comparison between the baseline and the proposed system. Dev set results are MTWVs, test set results are triplets of $_{l}A_{h}$ where $A$ is the ATWV, and $l$ and $h$ are the 2.5th and 97.5th percentile ATWV estimate respectively.
    }
    \centering
    \begin{tabular}{c cc ccc ccc c|cc ccc ccc ccc}
    \toprule
          Dataset && \multicolumn{2}{c}{LS-clean} && \multicolumn{2}{c}{LS-other} && \multicolumn{2}{c}{BNTR} & \multicolumn{2}{c}{AUD-NMI}\\
         KWS Feature & AUD Feature& Dev & Test && Dev & Test && Dev & Test & LS & BNTR \\
         \midrule
         MFCC&-&38.4&$_{38.7}$40.4$_{42.0}$&&16.2&$_{13.0}$14.3$_{15.7}$&&67.0 & $_{69.4}$70.7$_{71.9}$& - & -\\
         MFCC&MFCC&44.2&$_{44.0}$45.7$_{47.4}$&&21.1&$_{18.8}$20.3$_{21.9}$&&74.5&$_{76.4}$77.6$_{78.6}$&34.2&27.2\\
         MFCC&XLS-R&56.3&$_{54.9}$56.5$_{58.2}$&&30.9&$_{27.1}$28.7$_{30.4}$&&78.2&$_{80.4}$81.4$_{82.3}$&52.9&41.3\\
         \midrule
         XLS-R&-&73.2&$_{71.0}$72.7$_{74.4}$&&62.2&$_{61.6}$63.3$_{65.0}$&&84.8&$_{85.3}$86.1$_{86.9}$&-&-\\
         XLS-R&XLS-R&75.8&$_{74.7}$76.2$_{77.6}$&&65.5&$_{64.9}$66.4$_{67.9}$&&84.0&$_{84.7}$85.5$_{86.2}$&52.9&41.3\\
         \bottomrule
    \end{tabular}
    \label{tab:main}
\end{table*}

\subsubsection{Metrics}
\noindent \textbf{Term Weighted Value}: In our experiments, we report the term weighted values (TWV)~\cite{Fiscus2007,wegmann2013tao}, which is a measure of weighted recall and precision averaged across queries.
The TWV of a system on a set of queries, $\mathcal{Q}$, at a threshold, $\zeta$, is defined as:
\begin{align}
    \text{TWV} = 100 \times \bigl(1 - \frac{1}{\mathcal{Q}} \sum_{q \in \mathcal{Q}} (P_{miss}(q, \zeta) + \beta P_{FA}(q, \zeta)) \bigr),
    \label{eqn:twv}
\end{align}
where $P_{miss}(q, \zeta)$ is the probability of misses, $P_{FA}(q, \zeta)$ is the probability of false alarms and $\beta$ is a parameter controlling the relative importance of the two.
Following prior NIST evaluations~\cite{Fiscus2007}, we set $\beta=999.9$.
The threshold $~\zeta$ is tuned on the dev sets.
We report the maximum term weighted value (MTWV)---the TWV at the threshold which maximizes it---for the dev sets,
and the actual term weighted value (ATWV)---computed by using the threshold tuned on the corresponding dev set---for the test sets.
We adopt keyword-specific thresholding for across-query score normalization~\cite{miller2007rapid} in order to allow various queries with different score distributions to be compared with a single global threshold.

\noindent \textbf{Normalized Mutual Information}: We also report normalized mutual information (NMI) for the AUD systems in order to see how KWS performance correlates with the intrinsic quality of the AUD system used for pretraining.
NMI is computed by normalizing the mutual information between discovered units $\mathcal{U}$ and reference phones $\mathcal{P}$ by the sum of their entropies:
\begin{align}
    \text{NMI}(\mathcal{P}, \mathcal{U}) = 200 \times \frac{I(\mathcal{P}; \mathcal{U})}{H(\mathcal{P}) + H(\mathcal{U})} \%.
\end{align}
NMI takes values in $[0, 100]$, with 0 denoting completely uncorrelated units and 100 denoting perfect match.

\subsubsection{Model configuration and hyper-parameters}
We base the architecture of our model on~\cite{yusuf2023end}\footnote{Code available at: https://github.com/bolajiy/golden-retriever}.
The query encoder is a network with a 32-dimensional embedding layer for computing vector representations of each input grapheme, followed by 2 bidirectional GRU layers with 256 output units per direction per layer, and a 400-dimensional output projection layer whose outputs are summed along the sequence dimension to obtain the vectoral query representation.

The document encoder has 6 BLSTM layers with 512-dimensional output per direction per layer, followed by a 400-dimensional output layer.
We apply dropout of 0.4 between successive BLSTM layers, and down-sample by a factor of 2 after the fourth BLSTM layer.
This results in document encodings with frame durations of 40ms for XLS-R features and 20ms for MFCC.

The H-SHMM use 3-state HMM-GMMs with 4 diagonal-covariance Gaussians per HMM state, Dirichlet process with truncation parameter of 1, 100-dimensional unit embeddings ($\Matrix{\eta}^{l, u}$) and 5-dimensional language embeddings ($\Matrix{\alpha}^l$).

\subsubsection{Training details}
The neural networks are trained with the Adam optimizer~\cite{kingma2014adam}.
For pretraining, we use a cosine decay schedule with peak learning rate of 5e-4 and final learning rate of 1e-7, and train for 200k steps using mini-batch size of 256, with 10k warmup steps.
For finetuning (and the baseline with no pretraining), we train with the same step-based learning rate scheduling as~\cite{yusuf2023end} using mini-batch size of 32.
This starts with a learning rate of 0.002, halved whenever the validation loss (computed over 10\% of the training queries) does not improve over 4 epochs.
The training is stopped when validation loss does not improve for 10 epochs.
\subsection{Results}
Table~\ref{tab:main} shows the term weighted values of KWS with and without the proposed pretraining scheme, as well as the intrinsic AUD metrics.

When MFCC are used as input features to KWS, we find that pretraining with AUD learned pseudo-queries leads to significant improvements across all dev and test sets---with 5.3, 6.0 and 6.9 ATWV improvements respectively on the Librispeech-clean, Librispeech-other and BNTR test sets.
Furthermore, using XLS-R features for AUD (improving the quality of the learned acoustic units by a considerable margin) leads further significant ATWV improvements when compared to pretraining with MFCC-based acoustic units.

Replacing the KWS input with XLS-R unsurprisingly results in a much better performance, even compared to the pretrained MFCC-based systems, especially on the acoustically difficult Librispeech-other sets.
Furthermore, when we pretrain the XLS-R based KWS system with AUD labels, we observe +3.5 and +3.1 ATWV respectively on the Librispeech test-clean and test-other sets and no improvement on the BNTR.

\section{Conclusions}
In this paper, we have proposed a pretraining scheme for end-to-end keyword search.
Our approach uses acoustic unit discovery to label untranscribed data and construct pseudo-queries used to pretrain the KWS model, before finetuning on a small trasncribed dataset.
Our experiments show that pretraining can significantly improve KWS performance.

We envision future work doing larger scale and multilingual pretraining with acoustic unit targets in conjunction with or in place of transcribed data.
Another direction is to explore using more sophisticated word segmentation methods instead of our naive use of all n-grams of acoustic units.

\section{Acknowledgements}
The work was supported by Czech Ministry of Interior projects Nos. VJ01010108 ``ROZKAZ" and  by European Union’s Horizon Europe project No. SEP-210943216 ``ELOQUENCE".
Computing on IT4I supercomputer was supported by the Ministry of Education, Youth and Sports of the Czech Republic through e-INFRA CZ (ID:90254).
Computing on the ROYAL compute server was supported by the Turkish Directorate of Strategy and Budget under the ROYAL Project (CB SBB 2019K12-149250).



\bibliographystyle{IEEEtran}
\bibliography{mybib}

\end{document}